\documentclass[aps,prd,12pt,nofootinbib,,preprintnumbers %showpacs
,floatfix]{revtex4}
\usepackage[dvips]{graphicx}
\unitlength=1.2mm \preprint{JLAB-THY-07-720}
\begin{document}
\newcommand{\tr}{\mbox{tr}\,}
\newcommand{\Dslash}{{\mathchoice
    {\Dslsh \displaystyle}%
    {\Dslsh \textstyle}%
    {\Dslsh \scriptstyle}%
    {\Dslsh \scriptscriptstyle}}}
\newcommand{\Dslsh}[1]{\ooalign{\(\hfill#1/\hfill\)\crcr\(#1D\)}}
\newcommand{\leftvec}[1]{\vect \leftarrow #1 \,}
\newcommand{\rightvec}[1]{\vect \rightarrow #1 \:}
\renewcommand{\vec}[1]{\vect \rightarrow #1 \:}
\newcommand{\vect}[3]{{\mathchoice
    {\vecto \displaystyle \scriptstyle #1 #2 #3}%
    {\vecto \textstyle \scriptstyle #1 #2 #3}%
    {\vecto \scriptstyle \scriptscriptstyle #1 #2 #3}%
    {\vecto \scriptscriptstyle \scriptscriptstyle #1 #2 #3}}}
\newcommand{\vecto}[5]{\!\stackrel{{}_{{}_{#5#2#3}}}{#1#4}\!}
\newcommand{\vdot}{\!\cdot\!}

\bibliographystyle{apsrev}

\title{\large Lattice QCD Beyond Ground States}
\author{Huey-Wen Lin\footnote{Speaker}}
\email{hwlin@jlab.org}
\affiliation{Thomas Jefferson National Accelerator Facility,
Newport News, VA 23606}
\author{Saul D. Cohen}
\email{sdcohen@jlab.org}
\affiliation{Thomas Jefferson National Accelerator Facility,
Newport News, VA 23606}
\date{September 2007}

\begin{abstract}
In this work, we apply black box methods (methods not requiring input) to find excited-state energies. A variety of such methods for lattice QCD were introduced at the 3rd iteration of the numerical workshop series. We first review a selection of approaches that have been used in lattice calculations to determine multiple energy states: multiple correlator fits, the variational method and Bayesian fitting. In the second half, we will focus on a black box method, the multi-effective mass. We demonstrate the approach on anisotropic lattice data, extracting multiple states from single correlators. Without complicated operator construction or specialized fitting programs, the black box method shows good consistency with the traditional approaches.
\end{abstract}

\maketitle

\clearpage
\section{Introduction}

Lattice QCD has successfully provided many experimental quantities from first-principles calculation, sometimes predicting experimental results before they are measured. However, its success has mostly been restricted to measurements of ground-state quantities. In spectroscopy, for example, there are many poorly known states which require theoretical input to be identified. At the EBAC
%\cite{beac-link}
at Jefferson Lab, dynamical reaction models have been developed to interpret extracted $N^*$ parameters in terms of QCD\cite{Lee:2006xu,Matsuyama:2006rp}.

Among these excited nucleon states, the nature of the Roper resonance, $N(1440)$ or $N^\prime$, has been the subject of interest since its discovery in the 1960s.
%\cite{FIXME}.
It is quite surprising that the nucleon's excited-state mass is lower than its negative-parity partner, a phenomenon never observed in meson systems. There are several beyond-the-quark model interpretations of the Roper state, for example, as the hybrid state that couples predominantly to QCD currents with some gluonic
contribution\cite{Carlson:1991tg} %,Li:1991yb}
or as a five-quark (meson-baryon) state\cite{Krehl:1999km}.
Earlier quenched lattice QCD calculations~\cite{Sasaki:2001nf,Mathur:2003zf,Guadagnoli:2004wm,Leinweber:2004it, Sasaki:2005ap,Sasaki:2005ug,Burch:2006cc}, found a spectrum inverted with respect to experiment, with the $P_{11}$ $N^\prime$ heavier than the opposite-parity state $S_{11}$. However, this has been clarified by the study in Ref.~\cite{Mathur:2003zf} with a larger lattice box $\approx (3.2\mbox{ fm})^3$ and a broad pion mass study from 870~MeV to 180~MeV. They found that the mass of the $N^\prime$ becomes lighter than the $S_{11}$ when the pion mass gets lighter than 300--400~MeV. The findings of Ref.~\cite{Mathur:2003zf} indicate a rapid crossover of the first positive- and negative-parity excited nucleon states close to the chiral limit.
We need a full dynamical lattice QCD calculation with proper extrapolation to (or calculation nearby) the physical pion mass to resolve the Roper issue once and for all.

\begin{figure}\label{fig:roper_data}
\includegraphics[width=\textwidth]{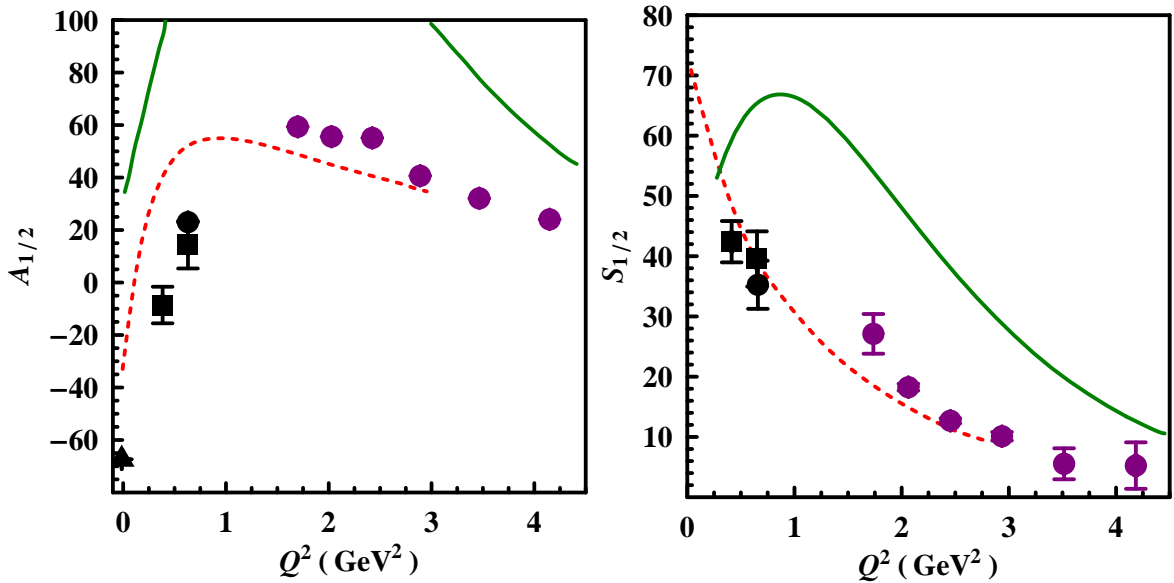}
\caption{Preliminary results on the transverse (left) and longitudinal (right) amplitudes of the Roper resonance. The solid lines represent non-relativistic quark model predictions, the dashed lines are predictions of relativistic quark models.}
\end{figure}

The excited transition form factor is also an interesting quantity for lattice QCD to explore. Experiments at Jefferson Laboratory, MIT-Bates, LEGS, Mainz, Bonn, GRAAL, and Spring-8 offer new opportunities to understand in detail how nucleon resonance ($N^*$) properties emerge from non-perturbative aspects of QCD. Consider the Roper transition form factor, for example. Figure~\ref{fig:roper_data} shows preliminary data from CLAS collaboration that covers a large range of photon virtuality $Q^2$. In the region $Q^2 \le 1.5 \mbox{ GeV}^2$, both the transverse amplitude $A_{1/2}(Q^2)$, and the longitudinal amplitude $S_{1/2}(Q^2)$, drop rapidly in magnitude. This is well described in relativistic quark models with light-cone dynamics, and the sign is consistent with the non-relativistic version. However, in the low-$Q^2$ region, $A_{1/2}(Q^2)$ becomes negative; this is not understood within constituent quark models and requires inclusion of meson degrees of freedom.
There are various QCD-based hadron models such as the well developed constituent quark model\cite{Capstick:2000qj,Aznauryan:2007ja} and the covariant model based on Dyson-Schwinger Equations\cite{Maris:2003vk}.
A model-independent study of these quantities from lattice QCD will serve as valuable help to phenomenologists in analyzing experimental data from stronger theoretical ground to gain a better understanding of low-$Q^2$ physics. Recently, we have submitted a proposal for studying this physics to the USQCD collaboration; we expect the first lattice QCD calculation of this quantity fairly soon.

Even when one only cares about getting ground-state quantities, one cannot quote precise numbers without estimating excited-state contamination. By explicitly adding excited-state degrees of freedom, we can obtain more accurate assessment of ground-state quantities. Examples will be given in later sections.

The structure of this proceeding is as follows: Sec.~\ref{sec:Lqcd} gives a brief lattice QCD introduction for the non-latticists. We discuss various approaches that have been used in lattice QCD for extracting excited states, such as multiple-correlator least-squares fitting, the variational method\cite{Michael:1985ne,Luscher:1990ck}, Bayesian fitting\cite{Michael:1993yj,Michael:1994sz,Lepage:2001ym}, and last but not least, black box methods\cite{Fleming:2004hs}, including both analytic excited effective masses and linear predictions. Conclusions and future outlook are given in Sec.~\ref{sec:Summary}.

\section{Lattice QCD}\label{sec:Lqcd}
This section provides a brief introduction to lattice QCD calculations for non-experts.

Quantum chromodynamics (QCD), a non-Abelian (Yang-Mills) gauge theory with gauge group $SU(3)$, is currently the best description of the strong interaction. In this theory, the physical observables are calculated from the path integral
\begin{eqnarray}\label{eq:PathInt}
\langle \Omega | O |\Omega \rangle & = &
  \frac{1}{Z} \int [dU][d\psi][d\overline{\psi}]
              e^{-S_F(U, \psi, \overline{\psi})-S_G(U)}O(U, \psi,\overline\psi)
\end{eqnarray}
in Euclidean spacetime.
% THIS
%At weak coupling, one can calculate this integral analytically in
%perturbation theory. However, when one reaches strong coupling, the
%series no longer converges well. Such calculations call for for
%nonperturbative treatment.
%% OR
The asymptotic freedom of QCD is a consequence of the negativity of the beta function for small numbers of flavors or large numbers of colors in non-Abelian gauge theory. When the coupling becomes weak at short distances, we can perform perturbative calculations to extract physics analytically. However, at large distances the strong coupling which accounts for quark confinement leads to a perturbative series that no longer converges, and we need a non-perturbative treatment, like lattice QCD, to extract physics from QCD.

Lattice QCD is a discrete version of continuum QCD theory. The path integral over field strengths at infinitely many spacetime points from continuum QCD is approximated using only finitely many points in a discrete Euclidian space lattice with periodic boundary conditions. In momentum-space, this automatically provides both an ultraviolet cutoff at the inverse lattice spacing and an infrared cutoff at the inverse box size.

By discretizing spacetime, we can then use Monte Carlo integration combined with the ``importance sampling'' technique to calculate the path integral numerically. At the end of the day, we need to take $a \rightarrow 0$ and $V \rightarrow \infty$ to reach the continuum limit. However, to simulate at the real pion mass (while at the same time keeping the lattice box big enough to avoid finite-volume effects) would require much faster supercomputers than are currently available to the community. Thus, we normally calculate with a few unrealistically large values of the pion mass and then use low-energy chiral effective theory to extrapolate to the physical pion mass.

%%% 2pt and 3pt

Most observables calculated in lattice QCD can be categorized as two-point and three-point Green functions:
\begin{eqnarray}
C_{\rm 2pt}^{\Gamma}(X_{\rm src}, X_{})
&=& \sum_{\alpha,\beta} \Gamma^{\alpha,\beta} \langle   J
(X_{}) J (X_{\rm src}) \rangle_{\alpha,\beta}\\
C_{\rm 3pt}^{\Gamma}(X_{\rm src}, X_{\rm }, X_{\rm snk})
&=& \sum_{\alpha,\beta} \Gamma^{\alpha,\beta} \langle   J (X_{\rm
snk}) O( X_{\rm}) J(X_{\rm src}) \rangle_{\alpha,\beta}
\end{eqnarray}
After taking momentum (and spin for baryons) projection (ignoring the variety of boundary condition choices possible), these become
\begin{eqnarray}
C_{\rm 2pt}(t)&=& \sum_n Z_n(p) e^{-E_n t},\label{eq:two-pt}\\
C_{\rm 3pt}(t_i,t,t_f,\vec{p}_i,\vec{p}_f)  &=&
\sum_n \sum_{n^\prime}  Z_{n^\prime}(p_f) Z_{n}(p_i)
     e^{-(t_f-t)E_n^\prime(\vec{p}_f)}e^{-(t-t_i)E_n(\vec{p}_i)}
 \times \mbox{form factors},\label{eq:excited_3pt}
\end{eqnarray}
where $Z$ contains not only the overlap factor between the vacuum and the operator, but also the kinetic factor with energies or masses.
Note that in either of the two- or three-point functions, the different states are mixed, and the signal decays exponentially as a function of $t$. This gives us a challenge: we want to disentangle the different channels, and this must be done before the signal dies out at large times.
%%%%%%%%

One usually obtains ground-state observables at large $t$, after the excited states die out, a method which requires large time dimension.
The lighter the particle is, the longer the $t$ required to get a ``good enough'' ground state. As lattice simulations go to light sea pion masses and finer lattice spacings, even getting an accurate ground state would not be easy without including excited states in the analysis.

Moreover, sometimes the large-$t$ solution does not work well, since the signal-to-noise ratio rapidly decreases. For example, when LHPC \& SESAM\cite{Dolgov:2002zm} calculated the quark helicity distribution, they found a 50\% increase in error at $t_{\rm sep} = 14$ over the error at separation 12. Confronting the excited states might be a better strategy than avoiding them, especially for precision calculations in the future.

\section{Probing Excited States}\label{sec:ExcitedStates}
%In this section, I am going to emphasize on how does one obtain the
%spectrum from lattice calculations.
There are various approaches that have been used widely by lattice theorists to probe excited states: multiple-correlator fits, variational methods and Bayesian fitting methods.
%(see the workshop proceedings from G.~Fleming and P.~Petreczky).
We would particularly like to continue the discussion of black box methods and show how they may be implemented on lattice QCD data for the first time.

Note that the data on which we demonstrate these methods is from a ``quenched'' (that is, ignoring fermion loops in the sea) study on $16^3\times64$ anisotropic lattices with Wilson gauge action and nonperturbative clover fermion action. The spatial lattice spacing $a_s$ is about 0.125~fm with anisotropy 3 (that is, temporal spacing $a_t^{-1} \approx 6$~GeV). Specifically, our data are proton correlators using 7 gaussian smearing parameters, 0.5--6.5 in steps of 1.0, including both smeared-point and smeared-smeared source-sink combinations.

\begin{figure}
\includegraphics[width=0.7\textwidth]{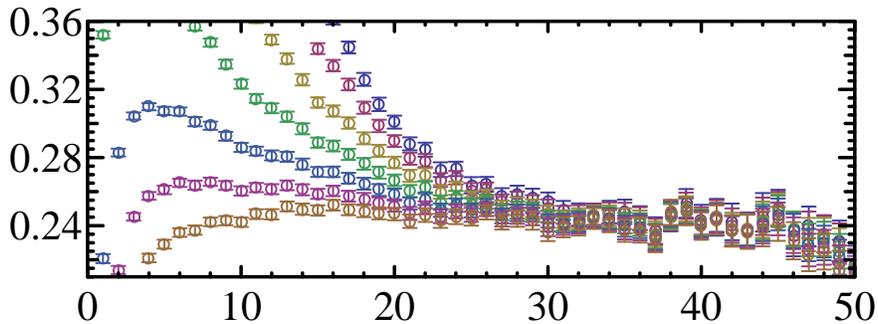}
\caption{Effective mass plots from the 7 smeared-point proton correlators used in this work}
\label{fig:meff}
\end{figure}

\subsection{Multiple-Correlator Fits}
The brute-force way of extracting excited states is to fit using a form that includes them explicitly. Usually, one needs multiple correlators as input to make the fit program converge. With multiple smeared correlators, one minimizes the quantity
%(FIXME: We probably want to consolidate this definition with the one I
%       use later in the Bayesian section. Change that one to a reference
%       to this one...)
\begin{eqnarray}
\chi^2=\sum_s \frac{(G_s(t)-\sum_n a_{n,s}e^{-E_nt})^2}
                   {\sigma_s^2(t)},\label{eq:chi2}
\end{eqnarray}
where the correlators $G_s$ are each fit by the sum of $n$ exponentials, having independent amplitudes but the same masses (energies) $E$. To extract $n$ states, one at needs at least $n$ distinguishable input correlators.

Using this technique, we find $E_0 = 0.240(3)$, $E_1 = 0.40(3)$ and $E_2 = 0.67(4)$.
%(FIXME: Extend the discussions to cover the difficulities, etc.)
%
%[FIXME] One in general needs multiple correlators to extract multiple states from

\subsection{Variational Method}
We construct an $r \times r$ spectrum correlation matrix, $C_{ij}(t)$, where each element of the matrix is a correlator composed from different smeared sources or operators ${\cal O}_i$ and ${\cal O}_j$. Then we consider the generalized eigenvalue problem
\begin{eqnarray}
C(t)\psi=\lambda(t,t_0)C(t_0)\psi,
\end{eqnarray}
where $t_0$ determines the range of validity of our extraction of the lowest $r$ eigenstates. If $t_0$ is too large, the highest-lying states will have exponentially decreased too far to have good signal-to-noise ratio; if $t_0$ is too small, many states above the $r$ we can determine will contaminate our extraction. Over some intermediate range in $t_0$, we should find consistent results.

If the eigenvector for this system is $|\alpha\rangle$, and $\alpha$ goes from 1 to $r$. Thus the correlation matrix can be approximated as
\begin{eqnarray}
C_{ij}=\sum_{n=1}^r v_i^{n*} v_j^n e^{-tE_n},
\end{eqnarray}
with modified eigenvalue
\begin{eqnarray}
\lambda_n(t,t_0)=e^{-(t-t_0)E_n},
\end{eqnarray}
by solving
\begin{eqnarray}
C(t_0)^{-1/2}C(t)C(t_0)^{-1/2}\psi=\lambda(t,t_0)\psi.
\end{eqnarray}

\begin{figure}
\includegraphics[width=0.7\textwidth]{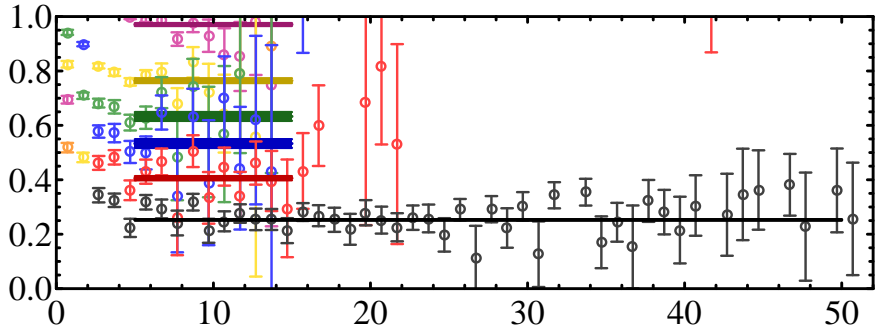}
\caption{Effective mass plots from generalized eigenvalues of a $7 \times 7$ smeared-smeared proton correlator matrix from anisotropic lattices.}
\label{fig:variational}
\end{figure}

Here we apply the above steps to a $7 \times 7$ smeared-smeared proton
correlator matrix. We solve the eigensystem for individual $\lambda_n$,
and use these eigenvalues to construct effective mass plots, as shown in
Figure~\ref{fig:variational}.
%Fit them individually with exponential form (red bars)
%Plotted along with effective masses
The masses of the lowest four eigenstates as fit in the regions shown in the figure are: $E_0 = 0.2526(24)$, $E_1 = 0.406(7)$, $E_2 = 0.533(16)$ and $E_3 = 0.633(16)$.

%The result is

\subsection{Bayesian Fitting}

One method for extracting additional masses from a single correlator is to apply more sophisticated fitting techniques. Using information we already know, we can introduce what are known as Bayesian priors to constrain our fits. These constraints may improve the precision with which difficult-to-determine masses are found by a fitting method.

The Bayesian technique applies known information to shift the estimated probabilities of correctness for parameter sets being considered by a fitting procedure. This prior information may be as simple as the assertion that all physical masses must be positive ($E_i>0$) or that each new excited state must be heavier than the states preceding it ($E_i > E_{i-1}$). In our case, we will attempt to extract states from a correlator in an iterative fashion. Assuming we have somehow acquired a ground state for our data (say, by a non-Bayesian fit), we can make the prior assumption that the ground state is actually equal to that fitted value within the determined uncertainty.

Using this prior information, we wish to modify the likelihood estimate for all parameter sets considered by our Bayesian fitter. The probability of correctness for a particular parameter set is generally represented by the $\chi^2$ value, which for a non-Bayesian fit is
\begin{equation}
\chi^2_{\rm dof} = \frac{1}{N-n} \sum_{i=1}^N
       \left(\frac{y_i - f(\{a,E\}_{0..n-1},x_i)}{\delta y_i}\right)^2,
\end{equation}
where we fit $N$ data points $y\pm\delta y$ to some form $f$ parametrized by $n$ amplitudes $A$ and masses (energies) $E$. Rather than allowing the fitter to arbitrarily change the amplitudes and masses, we would like it to take into consideration the masses we have already determined. Therefore, we simply add a term that will increase $\chi^2$ (decrease the probability of correctness) if the fitter moves away from these values:
\begin{equation}
\Delta \chi^2 = \sum_{j=0}^{m-1}
            \left(\frac{E_j-\hat{E}_j}{\delta \hat{E}_j}\right)^2,
\end{equation}
where we have already determined the first $m$ masses to be $\hat{E}\pm\delta \hat{E}$.

\begin{figure}
\includegraphics[width=0.7\textwidth]{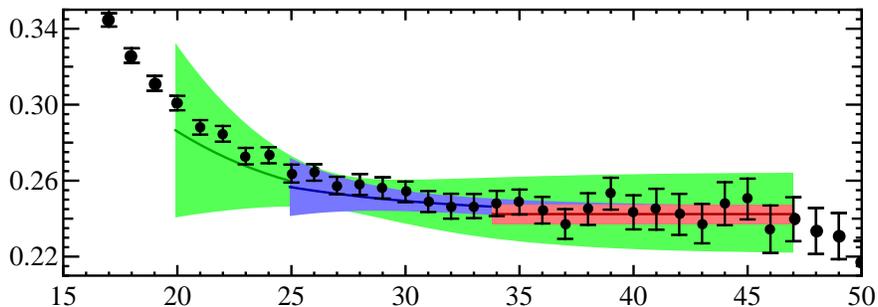}
\caption{A series of effective mass plots showing (red) the standard ground-state fit, (blue) an excited-state fit using the ground state as a Bayesian prior, and (green) the second-excited fit using ground and first-excited masses as priors.}
\label{fig:bayes}
\end{figure}

We apply such a Bayesian fit to the most pointlike of the correlators mentioned in the previous section. The results are shown in Figure~\ref{fig:bayes}. In the top-most plot, an ordinary fit determines the ground-state mass $E_0 = 0.243(5)$ over a short plateau $t \in [35,47]$. Using this prior information, we extend the fit region to $t \in [25,47]$, finding $E_0 = 0.2421(22)$ and $E_1 = 0.47(11)$. Finally, we apply both these masses to the fit region $t \in [20,47]$, finding $E_0 = 0.2420(8)$, $E_1 = 0.463(26)$ and $E_2 = 0.7(9)$. Since our second-excited mass is not statistically distinguishable from noise, it does not make sense to continue this procedure to yet-higher masses.

\subsection{Black Box Method}
In the 3rd iteration of this workshop, G.~T.~Fleming\cite{Fleming:2004hs} talked about ``black box methods'' used in NMR (nuclear magnetic resonance) imaging, where practitioners deal with problems of the form
\begin{eqnarray}
y_n = \sum_{k=1}^K a_k e^{i\phi_k} e^{(-d_k+i2\pi f_k)t_n} + e_n.
\end{eqnarray}
This is similar to lattice correlators, which have the general form
\begin{eqnarray}
y_n=\sum_{n=1}^K a_n e^{-E_n t}=\sum_{n=1}^K a_n \alpha_n^t,
\end{eqnarray}
where $\alpha_n=e^{-E_n}$. This forms a Vandermonde system
\begin{eqnarray}
y = \Phi a, ;\ \mbox{where }  \Phi  =\left(
\begin{array}{ccccc}
\phi_1(t_1,\alpha)    & \cdots  & \phi_K(t_1,\alpha) \\
\vdots                &  \ddots & \vdots  \\
\phi_1(t_N,\alpha)    & \cdots  & \phi_K(t_N,\alpha)  \\
\end{array}
\right).
\end{eqnarray}

Given a single correlator (with sufficient length of $t$), one can in principle solve for multiple $a$'s and $\alpha$'s
\begin{eqnarray}
\left(
\begin{array}{c}
y_1\\
y_2\\
\vdots \\
y_N
\end{array}
\right) =\left(
\begin{array}{ccccc}
\alpha_1     & \alpha_2  & \cdots & \alpha_K \\
\alpha_1^2   & \alpha_2^2 & \cdots & \alpha_K^2 \\
\vdots  & \cdots &\cdots & \vdots \\
\alpha_1^N & \alpha_2^N & \cdots & \alpha_K^N  \\
\end{array}
\right) \times \left(
\begin{array}{ccccc}
a_1 \\
a_2 \\
\vdots \\
a_K
\end{array}
\right)
\end{eqnarray}
The question is: can one solve for $N/2$ states from one correlator of length $N$?

\subsubsection{Analytic approach}
In the $K=1$ case, the system dominated by the ground state,
\begin{eqnarray}
\left(
\begin{array}{c}
y_t\\
y_{t+1}\\
\end{array}
\right) =\left(
\begin{array}{ccccc}
\alpha_1^{t}      \\
\alpha_1^{t+1}     \\
\end{array}
\right) \times \left(
\begin{array}{ccccc}
a_1
\end{array}
\right)
\end{eqnarray}
The solution is easy $\alpha_1 = y_{t+1}/y_t$. This is familiar to lattice theorists as the ``effective mass'' $M_{\rm eff} = \log(y_{t+1}/y_t)$. Applying this formula at various times produces an effective mass plot, which is helpful in determining appropriate fitting ranges, away from excited-state contamination.
%See the website {\tt http://www.jlab.org/~hwlin/} for animation demonstrating this technique.

Two states (the $K = 2$ case) may be extracted using
\begin{eqnarray}
\left(
\begin{array}{c}
y_1\\
y_2\\
y_3 \\
y_4
\end{array}
\right) =\left(
\begin{array}{ccccc}
\alpha_1     & \alpha_2 \\
\alpha_1^2   & \alpha_2^2  \\
\alpha_1^3   & \alpha_2^3 \\
\alpha_1^4 & \alpha_2^4  \\
\end{array}
\right) \times \left(
\begin{array}{ccccc}
a_1 \\
a_2
\end{array}
\right),
\end{eqnarray}
which leads to a more complicated solution
\begin{eqnarray}
\alpha_{1,2} = \frac{y_1 y_4-y_2 y_3 \pm \sqrt{(y_2y_3-y_1y_4)^2+4(y_2^2-y_1y_3)(y_2y_4-y_3^2)}}{2(y_1y_3-y_2^2)}.
\end{eqnarray}
One thing to note that is that the modified correlator approach, proposed by D.~Gaudagnol~et~al.\cite{Guadagnoli:2004wm}
is mathematically identical to the product of our solutions $\alpha_{1,2}$, giving the sum $E_0+E_1$.

%We first try this formulation on a toy model: consider three states with masses 0.5, 1.0, 1.5 and having the same amplitude, and adding in random Gaussian noise of various magnitudes.
%The results can be seen on the website: {\tt http://www.jlab.org/\~hwlin/}.

\begin{figure}
\includegraphics[width=\textwidth]{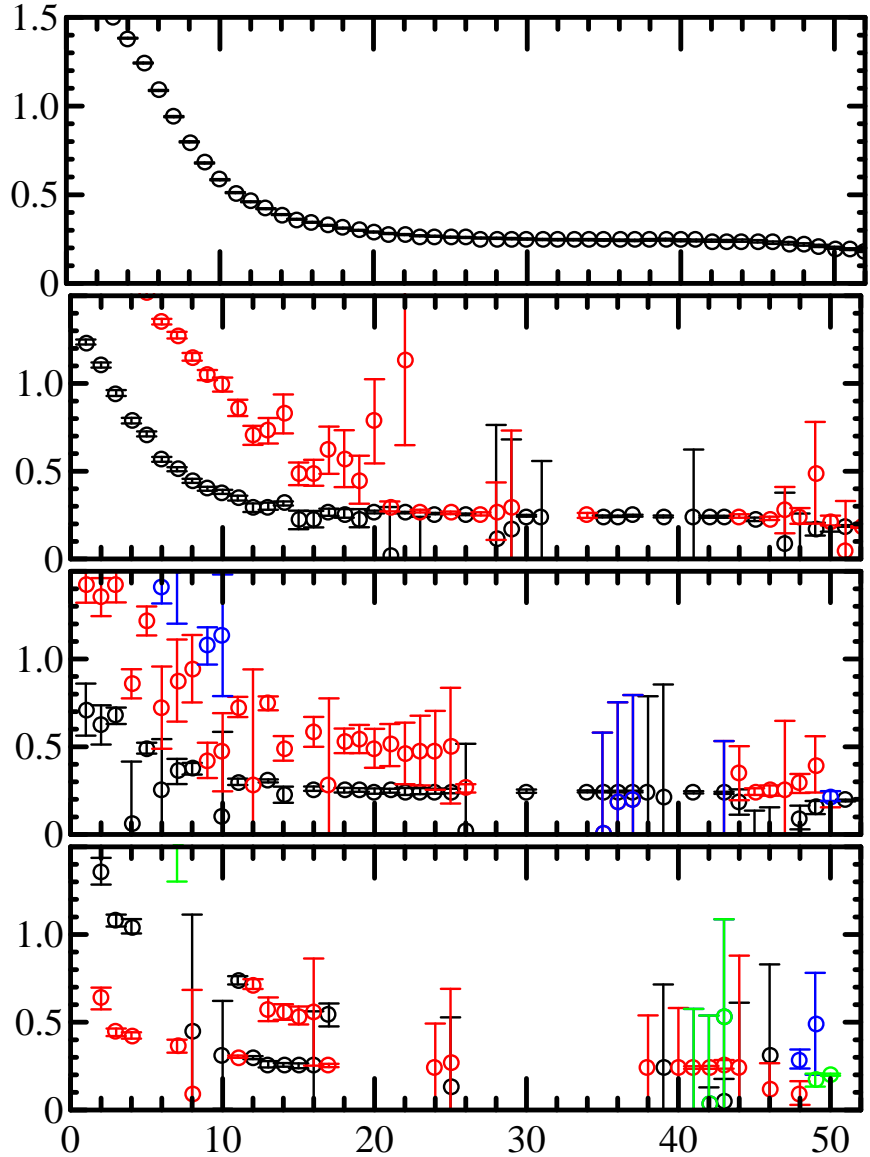}
\caption{Higher-effective mass plots with (top-to-bottom) one, two, three and four masses. The colors indicate the (black) ground, (red) first-excited, (blue) second-excited and (green) third-excited states.}
\label{fig:multieff}
\end{figure}

An example using our nucleon data is given in Fig.~\ref{fig:multieff}.

Such extensions of the effective mass technique provide simple but slightly different points of view from what we normally do in lattice calculations. Simple algebra gives us multiple states from a single correlator, which might be difficult to achieve using least-squares fitting. It is possible to calculate the second-excited state in similar fashion. However, there is a limitation to this analytic approach: Abel's Impossibility Theorem, which states that algebraic solutions are only possible for $N < 5$. To obtain more states, we would need to solve the resulting polynomial equations numerically.

\subsubsection{Linear prediction}
%(This part of the work is in collobaration with Saul D. Cohen.)
If we are interested in extracting $K$ states, we can first construct a polynomial with coefficients
\begin{eqnarray}
\Pi_{n=1}^K (\alpha -\alpha_n) = \sum_{n=1}^K p_n \alpha^{K-n}.
\end{eqnarray}
We can ``predict'' a data point at a later $t^\prime$ in terms of the earlier time points:
\begin{eqnarray}\label{eq:LP1}
y_{t^\prime} = -\sum_{n=1}^K p_n \alpha^{t^\prime-n} \mbox{ where } m \geq t^\prime
\end{eqnarray}
for ideal data. This is called a ``linear prediction''. However, for non-toy data,  background fluctuations will be present; thus, Eq.~\ref{eq:LP1} becomes
\begin{eqnarray}\label{eq:LP2}
y_{t^\prime} \approx -\sum_{n=1}^M p_n \alpha^{t^\prime-n} + v_{t^\prime}.
\end{eqnarray}
Therefore, to extract the multiple states, we need to solve the Vandermonde system
\begin{eqnarray}\label{eq:LP3}
\left(
\begin{array}{c}
y_M\\
y_{M+1}\\
\vdots \\
y_{N-1}
\end{array}
\right) = - \left(
\begin{array}{ccccc}
y_0     & \cdots & y_{M-1} \\
y_1     & \cdots & y_{M} \\
\vdots   & \ddots & \vdots \\
y_{N-M-1}     & \cdots & y_{N-2} \\
\end{array}
\right) \times \left(
\begin{array}{ccccc}
p_M\\
p_{M-1}\\
\vdots \\
p_{1}\\
\end{array}
\right) \mbox{ where } N \geq 2M.
\end{eqnarray}

\begin{figure}
\includegraphics[width=0.7\textwidth]{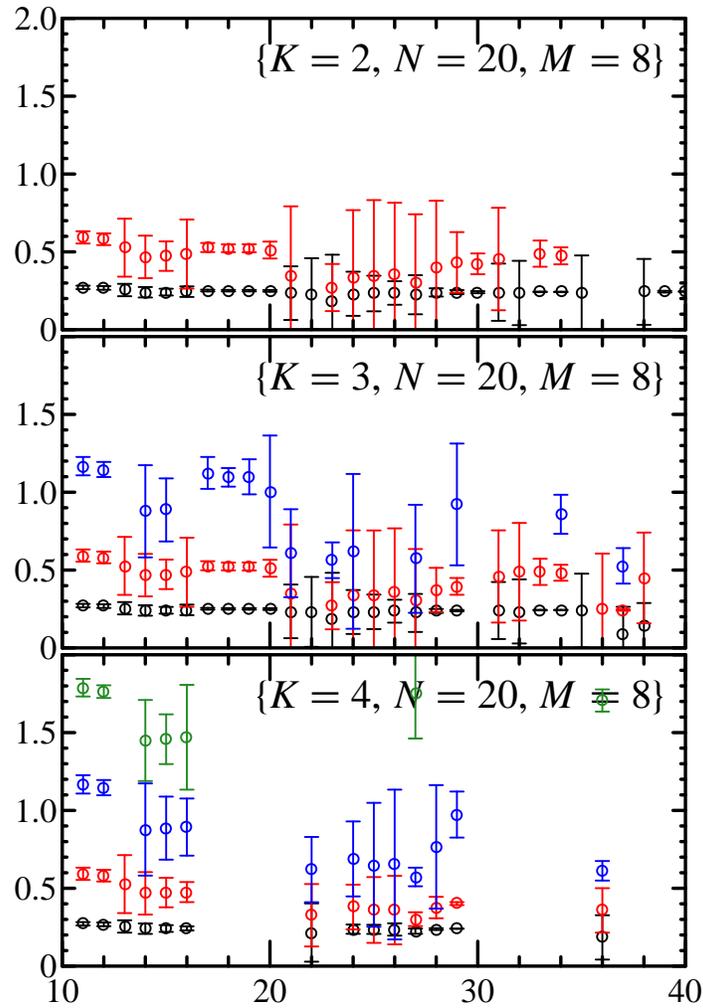}
\caption{Effective mass plots from the linear prediction black-box method at fixed parameters $N=20$ and $M=8$}
\label{fig:LPP-meff}
\end{figure}

We demonstrate this approach using the smallest smearing parameter (0.5) smeared-point correlator. There are a few parameters in the linear prediction method which we can tune: the number of desired states $K$, the number of time slices used to predict the later time point $N$, and the order of the polynomial $M$.
%
%We demonstrate how the results change by varying the parameters $M$ and %$N$ in the case $K=3$, as one can find in the following link: {\tt %http://www.jlab.org/~hwlin/}.
%
In this work, we will show a selection of the better choices in these degrees of freedom. Figure~\ref{fig:LPP-meff} shows the effective mass plot for $K=2,3,4$ from a single Gaussian smeared-point corrleator with fixed parameters $N=20$ and $M=8$. The excited states are consistent with each other as one increases the value of $K$. Since we have used a large value of $N$ to form the polynomial, each point uses information extracted from 20 time slices. Thus, one does not need a large plateau to determine the final mass. One also notes that since we only use a single correlator to extract multiple states, the multiple states will be correlated; that is, large errors on higher-excited states will make the ground state noisy as well. A future improvement would naturally be to extend this approach to multiple correlators.

\begin{figure}
\includegraphics[width=0.7\textwidth]{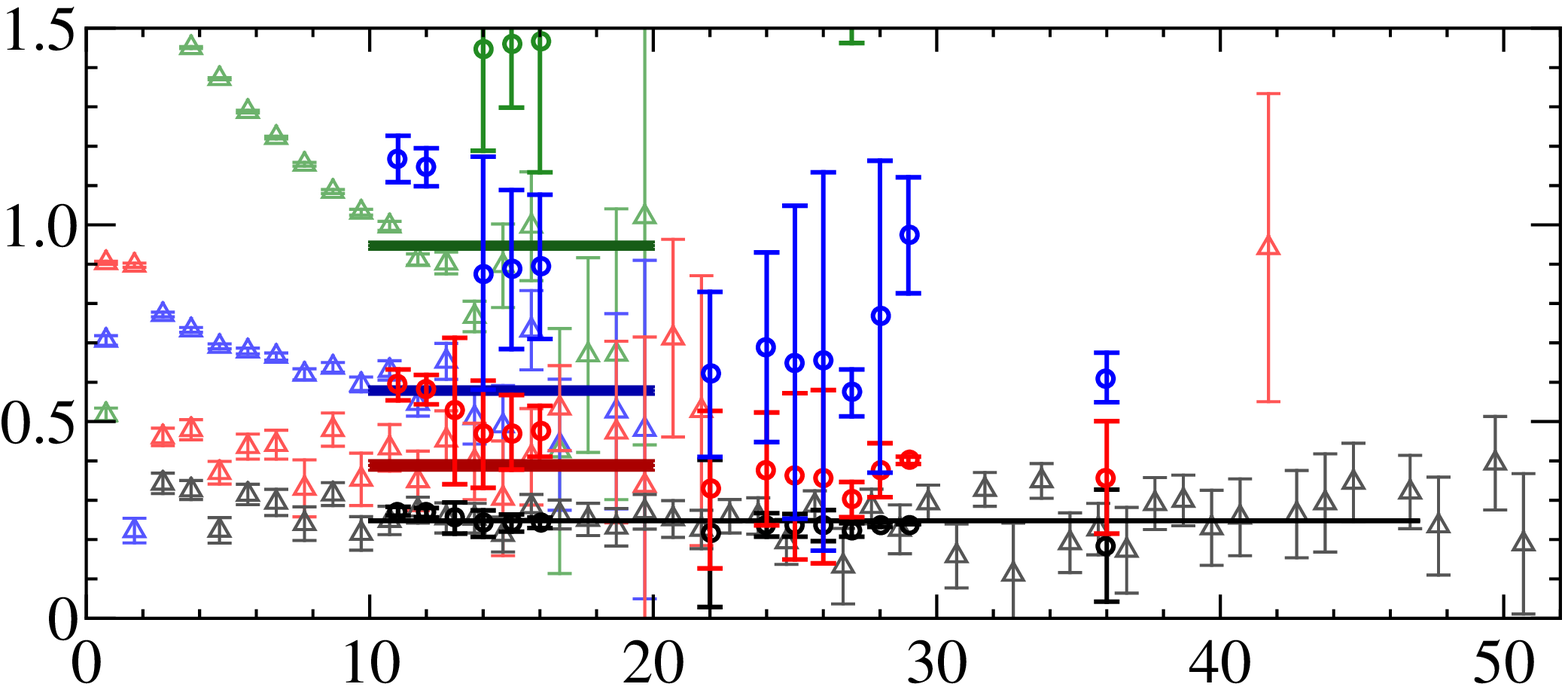}
\caption{Comparison between the variational method ($4 \times 4$) and linear prediction with 4 extracted states}
\label{fig:comp-meff}
\end{figure}

We compare this linear prediction approach with the variational method ($4 \times 4$, with smearing parameter ranging 0.5--3.5), as shown in Figure\ref{fig:comp-meff}. Here we shift time with respect to the linear prediction plot by 10 to have a better comparison with the plateau region from the variational method. For the ground state, the numbers are consistent with the result from the varitional approach, including the size of the errorbar. This is remarkable, given that the amount of input information is a factor of 16 less in the linear prediction approach. The first-excited state is consistent but has larger errorbar, which is no surprise. As for the second-excited state, it seems to be consistent with the varitional ones but definitely needs more stastictics. The third-excited state is much larger than expected from the varitional approach, which might be caused by contamination from even higher excited states.

\section{Conclusion and Outlook}\label{sec:Summary}
There is a lot of interesting physics involving excited states, but they are traditionally difficult to handle in lattice QCD. Even for physicists who are only interested in the ground states, the estimation of excited states can improve the precision of their calculations.

In this work, we adopt NMR-inspired methods which provide an interesting alternate point of view for looking at lattice QCD spectroscopy. We successfully extract multiple states from single correlators, working with quenched anisotropic lattice data. We also demonstrate various other methods for determining excited states (multiple-correlator fit, variational method and Bayesian fit). These methods require different amounts of input information and manual guidance, which may introduce systematics that are difficult to quantify. The black box methods provide alternatives without any ambigiuity and without the need for careful tuning.

Future work will extend this approach to multiple correlators\cite{Fleming:2007}, introducing more information to improve the signal.

\section*{ACKNOWLEDGMENTS}
HWL would like to thank the workshop organizer, George Fleming, for the workshop invitation and thus a possibility for the extension of earlier exploratory work, and SciDAC and Yale University for financial support to attend this workshop. Computations were performed on clusters at Jefferson Laboratory under the SciDAC initiative, using the Chroma software suite\cite{Edwards:2004sx} on clusters at Jefferson Laboratory using time awarded under the SciDAC Initiative. Authored by Jefferson Science Associates, LLC under U.S. DOE Contract No. DE-AC05-06OR23177. The U.S. Government retains a non-exclusive, paid-up, irrevocable, world-wide license to publish or reproduce this manuscript for U.S. Government purposes.
\vspace{-0.1in}

\clearpage
%%%%%%%%%%%%%%%%%%%%%%%%%%%%%%%%%%%%%%%%%%%%%%%%%%%%%%%%%%%%%%%%%%%%%%%%%%%%%%%%%%%%%%%%%%%%%%%%%%%%%%%%%%%%%%%%%%%%%%%%%
%\bibliography{actions}
\bibliography{ref}

\end{document}